\documentclass[doublecol,figures]{epl2}

\usepackage{ifthen}

\usepackage{graphicx}
\usepackage{color}
\usepackage{bm}
\usepackage{latexsym}

\newcommand{\mass}{\mathsf{m}}

\newcommand{\be}[1]{\begin{eqnarray}\ifthenelse{#1=-1}{\nonumber}{\ifthenelse{#1=0}{}{\label{e#1}}}}
\newcommand{\ee}{\end{eqnarray}} 

\newcommand{\hide}[1]{}
\newcommand{\tbox}[1]{\mbox{\tiny #1}}

\graphicspath{{figures/}}

\normalsize

\title{%
  Control of atomic currents using a quantum stirring device}

\author{%
  Moritz Hiller\inst{1,2,3}, Tsampikos Kottos\inst{1,2} \and Doron Cohen\inst{4}}

\institute{%
  \inst{1} Department of Physics, Wesleyan University, Middletown, Connecticut 06459, USA \\
  \inst{2} MPI for Dynamics and Self-Organization, Bunsenstra\ss e 10, D-37073 G\"ottingen, Germany \\
  \inst{3} Physikalisches Institut, Albert-Ludwigs-Universit\"at, Hermann-Herder-Str. 3, D-79104 Freiburg, Germany \\
  \inst{4} Department of Physics, Ben-Gurion University, Beer-Sheva 84105, Israel
}

\abstract{%
We propose a BEC stirring device which can be regarded as the
incorporation of a quantum pump into a closed circuit: 
it produces a DC circulating current in response to a cyclic adiabatic change 
of two control parameters of an optical trap. 
We demonstrate the feasibility of this concept and point out that 
such device can be utilized in order to probe the interatomic interactions.
}


\begin{document}
\maketitle


The realization of Bose-Einstein Condensation (BEC) of ultra-cold atoms in optical 
lattices and atom chips \cite{AK98} \hide{JBCGZ98,OTFYK01,GMEHB02,B05;FKCHMS00} 
and 
the availability of conveyor belts \cite{HHHR01} is considered 
to be a major breakthrough with potential applications in the arena of quantum 
information processing \cite{SFC02}, 
atom interferometry \cite{SHAWGBSK05,WABCDKPSSW05,ATMDKK97}, lasers 
\cite{AK98,MAKDTK97} and atom diodes and transistors \cite{MDJZ04,SKAH06,SAZ07}. A 
major advantage of BEC based devices, as compared to conventional solid-state structures, 
lies in the extraordinary degree of precision and control that is available, regarding 
not only the confining potential, but also the strength of the interatomic interactions, 
their preparation and the measurement of the atomic cloud. 
Accordingly it is envisioned that the emerging field of ``atomtronics" 
will provide a new generation of nanoscale devices. 
 

The possibility to induce DC currents by periodic (AC) modulation of the potential is 
familiar from the context of electronic devices. If an open geometry is concerned, it 
is known as ``quantum pumping" \cite{Thouless} \hide{NT84,BPT,Brouwer,AG99,SMCG99}, 
while for closed geometries we use the term ``quantum stirring" \cite{pml}.  
We consider below the stirring of condensed ultra-cold atoms \hide{\cite{MCWD00,RK99}}
due to the periodic variation of the on-site potentials 
and of the tunneling rates between adjunct confining traps. 
We show that the nature of the transport process depends crucially 
on the sign and on the strength of the interatomic interactions.
We distinguish between four regimes of dynamical behavior:
For strong repulsive interaction the particles 
are transported one-by-one, which we call {\em sequential crossing}; 
for weaker repulsive interaction we  
observe  either {\em gradual crossing} or coherent {\em mega crossing}; 
finally, for strong attractive interaction the particles  
{move together as one composite unit} from trap to trap. \\

\begin{figure}[t]
  \includegraphics[width=\columnwidth]{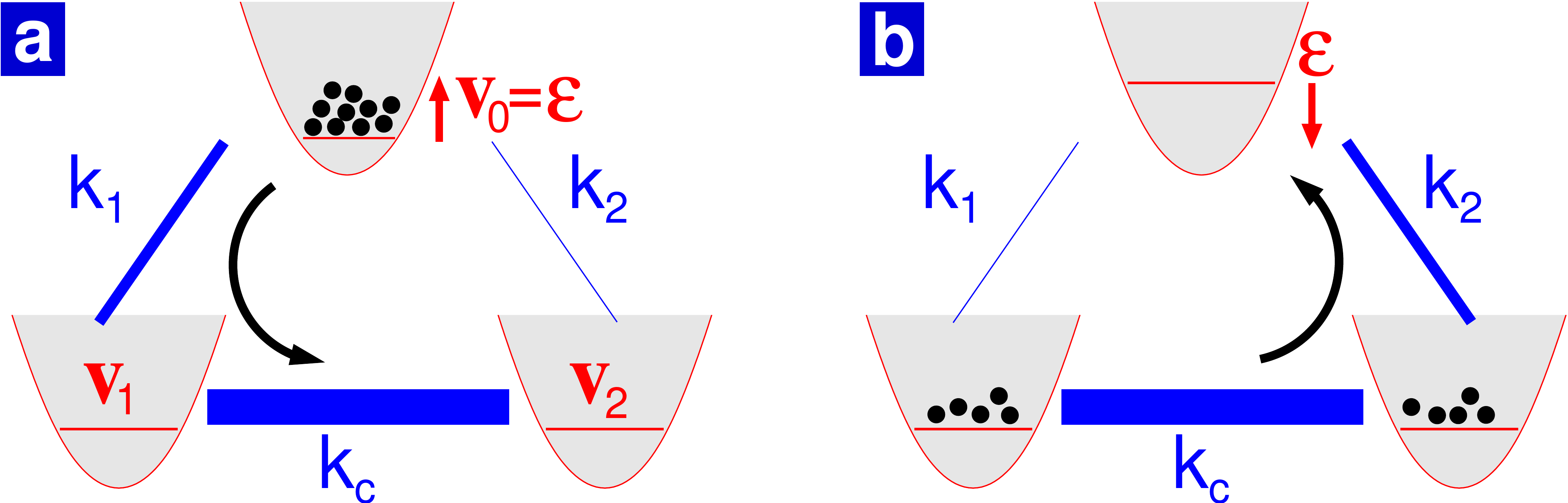}
  \caption{(Color online) 
    {\bfseries\sffamily Illustration of the model system.}
    In the first half of the cycle {\bfseries\sffamily a}
    the particles are transported from the {gated-well} to
    the {double-well} via the $k_1$ bond, while in the second half 
    of the cycle {\bfseries\sffamily b} the particles are transported 
    back from the {double-well} to the {gated-well} 
    via the $k_2$ bond. See the text for further details.
  }
\end{figure}


{\bf Model --}
The simplest model that captures the physics 
of quantum stirring is the three-site Bose-Hubbard 
Hamiltonian (BHH) \cite{BHK07,HKG06,SAZ07,FP03} (see Fig.~1) 
\footnote{
The three-site BHH is a prototype system for many recent studies \cite{HKG06}. 
A classical analysis of this system was performed in \cite{FP03} where
it was shown that for appropriate system parameters and 
initial conditions chaotic dynamics would emerge. In this work 
we consider adiabatic driving of ground state preparation and therefore 
chaotic motion is not an issue.}. 
\newcounter{noteA}
\setcounter{noteA}{\thefootnote}
We call the ${i{=}0}$ site ``{gated-well}" since 
we late assume that we have control over its potential
energy~${v_0=\varepsilon}$.  
The ${i{=}1{,}2}$ sites form a two level ``{double-well}" 
with potential energy ${v_1{=}v_2{=}0}$.  
The $N$~boson BHH is: 
\hide{
\be{1}
\hat{\mathcal{H}}&=&\sum_{i=0}^2 v_i {\hat n}_i+ \frac{U}{2}\sum_{i=0}^{2}{\hat n}_i ({\hat n}_i-1) -
k_c(\hat{b}_{1}^{\dagger}\hat{b}_{2}+\hat{b}_{2}^{\dagger}\hat{b}_{1})\nonumber\\ 
&& -k_1(\hat{b}_{0}^{\dagger}\hat{b}_{1}+\hat{b}_{1}^{\dagger}\hat{b}_{0}) - 
k_2(\hat{b}_{0}^{\dagger}\hat{b}_{2}+\hat{b}_{2}^{\dagger}\hat{b}_{0})
\label{QMBHH3}
\ee
}
\be{1}
\hat{\mathcal{H}}=\!\!
\sum_{i=0}^2 \left[v_i {\hat n}_i {+} \frac{U}{2}{\hat n}_i ({\hat n}_i{-}1)\right] 
{-}k_c\hat{b}_{2}^{\dagger}\hat{b}_{1}
{-}\!\!\sum_{i=1,2}\!\! k_i\hat{b}_{i}^{\dagger}\hat{b}_{0} 
{+}\mbox{h.c.}
\label{QMBHH3}
\ee
{We set ${\hbar{=}1}$ which corresponds to measuring energies 
in units of frequency. Furthermore, without loss of generality we choose 
time units such that} ${k_c{=}1}$. 
Accordingly the two single particle levels of the {double-well} are $\varepsilon_{\pm}=\pm1$. 
The annihilation and creation operators $\hat{b}_i$ and $\hat{b}_i^{\dagger}$ obey 
the canonical commutation relations $[\hat{b}_i,\hat{b}_j^{\dagger}]=\delta_{i,j}$ 
while the operators ${\hat n}_i=\hat{b}_i^{\dagger}\hat{b}_i$ 
count the number of bosons at site~${i}$. 
The interaction strength between two atoms in a single site is given 
by $U=4 \pi a_sV_{\tbox{eff}}/\mass$ where $V_{\tbox{eff}}$ 
is the effective volume, $\mass$~is the atomic mass, 
and $a_s$ is the $s$-wave scattering length 
{which can be changed by applying an additional magnetic field} \cite{Legg01}.
{The on-site potentials $v_i$ as well as the coupling strengths $k_i$ are controlled by
changing the confining potential. The main assumption underlying the BHH 
is that the single-particle ground state wavefunctions are sufficiently localized 
at the sites, and that for the temperatures involved they are well separated 
in energy from the excited single-particle levels. Experimentally, such    
deep trapping potentials can accommodate several hundred particles\cite{MCWW97}.}

The couplings between the {gated-well} and the two ends of the {double-well} are $k_1$ and $k_2$. We 
assume that both are much smaller than $k_c$ (for the two-mode BEC dynamics see for
example \cite{JM06}).
It is convenient to define the two control parameters of the pumping 
as ${X_1 = (1/k_2){-}(1/k_1)}$ and ${X_2 = \varepsilon}$.
\hide{
\be{2}
X_1 = \left(\frac{1}{k_2}-\frac{1}{k_1}\right), \qquad
X_2 = \varepsilon
\ee
}
By periodic cycling of the parameters $(X_1,X_2)$ 
we can obtain a non-zero amount ($Q$) of transported atoms per cycle. 
The pumping cycle is illustrated in Figs.~1,2.
Initially all the particles are located in the {gated-well}
which has a sufficiently 
negative on-site potential energy ($X_2<0$).  
In the first half of the cycle the coupling is biased in favor of 
the $k_1$ route (${X_1>0}$) while $X_2$ is raised 
{until (say) the {gated-well} is empty}
\footnote{{Later we estimate the required $X_2$~variation 
in order to have all the particles transferred from 
the gated-well to the double well. We also analyze smaller 
pumping cycles (see Fig.~3b) for which only a fraction 
of particles gets through.}}. 
In the second half of the cycle the coupling 
is biased in favor of the $k_2$ route (${X_1<0}$), while $X_2$ is
lowered until the {gated-well} is full. 
Assuming~$U{=}0$, the {gated-well} is depopulated via the $k_1$ route 
into the lower energy level $\varepsilon_{-}$ 
during the first half of the cycle, and re-populated 
via the $k_2$ route during the second half of the cycle. 
Accordingly the net effect is to have a non-zero~$Q$. 
If we had a single particle in the system, 
the net effect would be to pump roughly one particle 
per cycle. If we have~$N$ non-interacting particles, 
the result of the same cycle is to pump 
roughly~$N$ particles per cycle
\footnote{
Through the driving cycle the total number of bosons remains constant. 
The energy is not a constant of motion, but in the adiabatic limit 
the system comes back to the same state at the end of each cycle.}.
\newcounter{noteB}
\setcounter{noteB}{\thefootnote}
We would like to know what is the actual result using 
a proper quantum mechanical calculation, 
and furthermore we would like to investigate what is the 
effect of the interatomic interaction~$U$ on the result.\\

\begin{figure}[t]
  \includegraphics[width=\hsize]{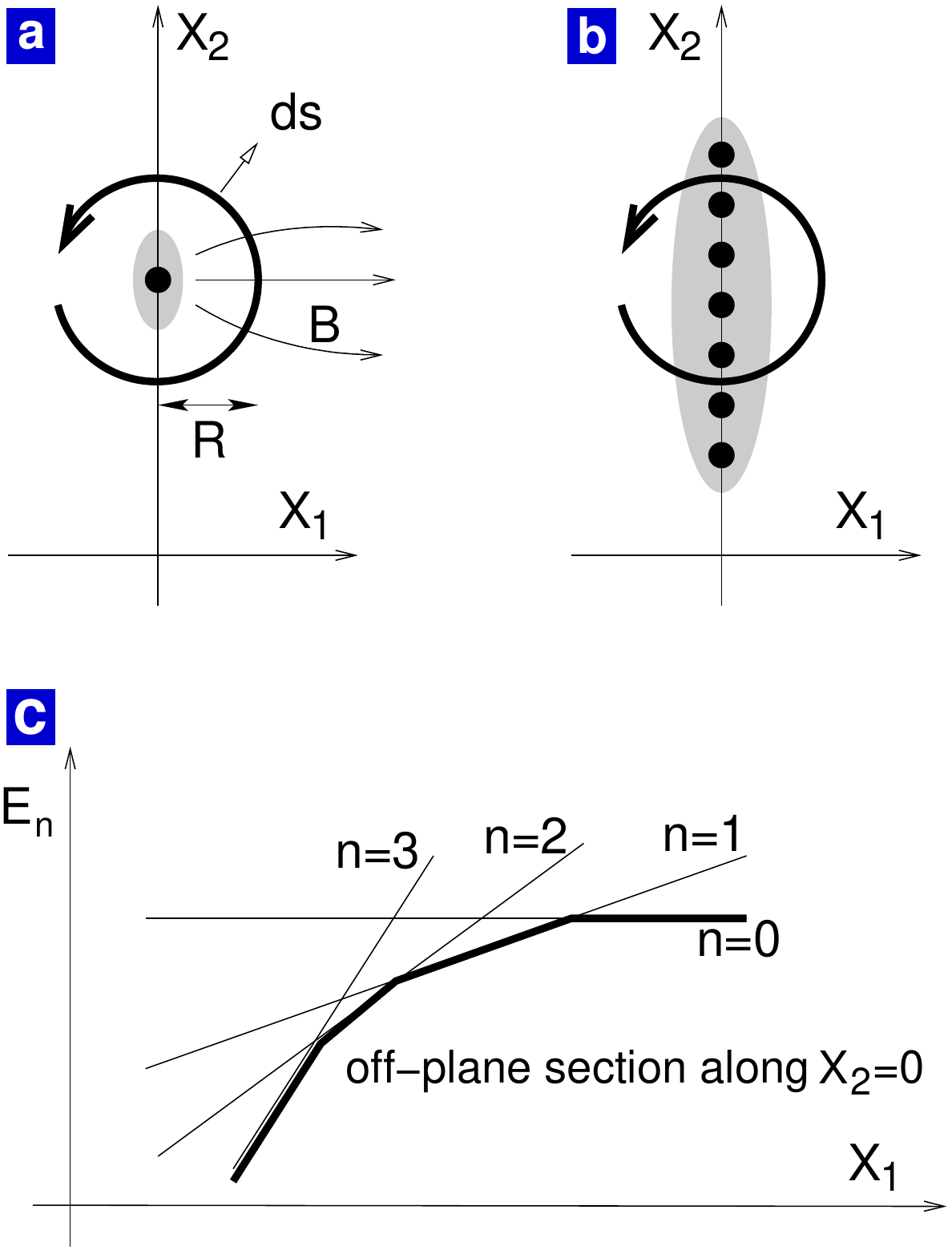}
  \caption{
    {\bfseries\sffamily Illustration of the pumping cycle.}
    See the text for further details.    
    For a large cycle that encircles the 
    whole shaded region we have $Q\approx N$.
    Position of the monopoles: {\bfseries\sffamily a} no interactions,
    {\bfseries\sffamily b} with interactions. 
    {\bfseries\sffamily c}, the energy levels along the ${X_1=0}$ axis
    are schematically plotted for an $N=3$ system.
  }
\end{figure}


{\bf Methods --}
{Within the framework of linear response theory 
the induced current is ${I=-G_1\dot{X}_1}$ if we change~$X_1$ 
and  ${I=-G_2\dot{X}_2}$ if we change~$X_2$. 
The coefficients $G_1$ and $G_2$ in these linear relations 
are {\em defined} as the elements of the geometric 
conductance matrix, and can be calculated using 
the Kubo formula approach (see below). 
Integrating the current over a full cycle we get}       
\be{3}
Q = \oint\limits_{\tbox{cycle}} Idt 
= -\oint (G_1 dX_1 + G_2 dX_2).
\ee     
In order to calculate the geometric conductance 
we use the Kubo formula approach 
to quantum pumping \cite{pmc} which is based on the 
theory of adiabatic processes \cite{Berry}\hide{Avron,Robbins}.
It turns out that in the strict adiabatic 
limit $G$ is related to the vector field $\bm{B}$ 
in the theory of the Berry phase {which is
known as the ``Berry Curvature''}. {The adiabatic slowness condition 
on $\dot{X}$ in the present context, taking 
into account the two-orbital approximation, 
is discussed in Section~4 of \cite{cnb}}.
Using the notations ${\bm{B}_1=-G_2}$ and ${\bm{B}_2=G_1}$ 
it is illuminating to rewrite Eq.(\ref{e3}) as 
\be{7}
Q = \oint \bm{B} \cdot d\vec{s} ,
\ee     
where we define the normal vector $d\vec{s}=(dX_2,-dX_1)$ 
as illustrated in Fig.~2. 
The calculation of the so-called Kubo-Berry Curvature is 
done using the following formula \cite{pmc}:   
\be{8}
\bm{B}_j=\sum_{n\ne n_0}
\frac{2 \ {\rm Im}[\mathcal{I}_{n_0n}] \ \mathcal{F}^j_{nn_0}}
{(E_n-E_{n_0})^2}.
\ee
Above $\mathcal{I}=i/2\left[k_1(\hat{b}_{0}^{\dagger}\hat{b}_{1}{-}\hat{b}_{1}^{\dagger}\hat{b}_{0})+ k_2
(\hat{b}_{2}^{\dagger}\hat{b}_{0}{-}\hat{b}_{0}^{\dagger}\hat{b}_{2})\right]$ 
is the averaged current
\footnote{{Due to the continuity equation the same result 
for $Q$ is obtained irrespective of whether we measure the current via 
the ${0{\mapsto}1}$ bond or via the ${2{\mapsto}0}$ bond. The advantage 
of using the the ``symmetrized" version for $\mathcal{I}$ is that the same amount
of particles is being transported during both halves of the cycle, 
allowing us to focus on (say) the first half and then double the result.}}
via the bonds $0{\mapsto}1$ and $2{\mapsto}0$,  
while $\mathcal{F}^j=-\partial \mathcal{H}/\partial X_j$ 
is the generalized force associated with the control parameter $X_j$.
{The index $n$ labels the eigenstates of the many-body 
Hamiltonian.} We assume from now on that $n_0$ 
is the BEC ground state\footnotemark[\thenoteA].

The advantage of the above, so called ``geometric"  
point of view is in the intuition that it gives 
for the result: 
{Formally the field~$\bm{B}$ is like a 
projection of a fictitious magnetic field in an embedding 
three-dimensional $X$~space (the third coordinate~$X_3$ 
if formally defined such that $\mathcal{I}=-\partial \mathcal{H}/\partial X_3$).    
The flux of {this fictitious magnetic field through any out-of-plane}
surface which is enclosed by the pumping cycle 
gives the so-called Berry phase, while the line integral {over this
fictitious magnetic field, i.e.} Eq.(\ref{e7}) {\footnote{{In the $X_3{=}0$~plane the
third component of the fictitious magnetic field is zero due to the
time-reversal symmetry \cite{pmc}.}}}, gives~$Q$. As implied by inspection 
of Eq.(\ref{e8}) the sources of~$\bm{B}$ are located 
at points where the ground level $n_0$ has a degeneracy  
with the next level. A simple argument implies 
that this ``magnetic charge" is quantized like Dirac monopoles, 
else the Berry phase would be ill-defined.   
For details see \cite{pmc}.} 
In our model system for $U{=}0$ all the ``magnetic charge" 
is concentrated in one point. As the interaction~$U$ 
becomes larger {the $(N{+}1)$-fold degeneracy of the levels 
is lifted,} and this ``magnetic charge" disintegrates 
into $N$ elementary ``monopoles" (see Fig.~2). 
We further discuss the energy spectrum in the next section.  \\


{\bf Regimes --}
We define the average coupling as $\kappa=(k_1{+}k_2)/\sqrt{2}$. 
In the zeroth-order approximation $k_1$ and $k_2$ are neglected,  
and later we take them into account as a perturbation. 
For $\kappa{=}0$ the number ($n$) of particles 
in the {gated-well} becomes a good quantum
number {hence we can associate the level index
$n$ with the number of particles in the gated-well.}
Furthermore, we adopt a ``two-orbital approximation":  
{we assume that there is non-zero occupation only 
in the gated-well and in the lower double-well level,  
which is valid if $NU \ll k_c$. Note also that we 
assume an adiabatic process\footnotemark[\thenoteB], and accordingly non-adiabatic 
transitions to the higher orbitals can be safely neglected \cite{cnb}.}

\begin{figure}[t]
  \includegraphics[width=\linewidth,clip]{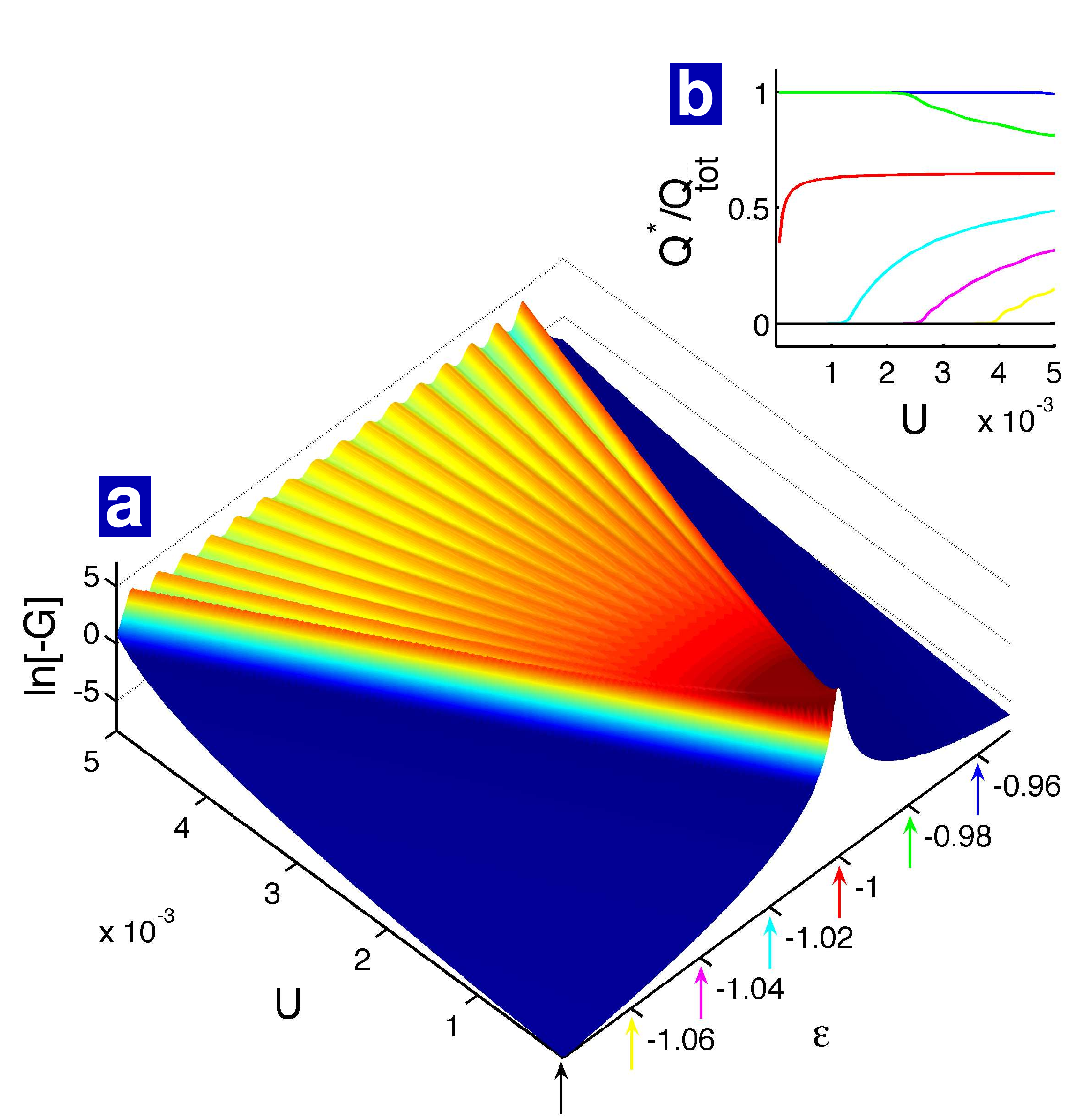}
  \caption{(Color online)
    {\bfseries\sffamily Conductance during the first
      half of the pumping cycle. a},  
    the conductance $G_2$ as a function 
    of the on-site potential~$\varepsilon$, 
    for various values of~$U$. In the numerical simulations (see also Fig. 4) we are 
    using exact diagonalization of the trimer Hamiltonian (\ref{QMBHH3}). For the 
    evaluation of $G$ we use Eq.~(\ref{e3}) while for $Q$ we use Eq.~(\ref{e8}).
    The other parameters are $N=16$ particles, 
    $\kappa=0.0003/\sqrt{2}$ and $(k_1-k_2)/\sqrt{2}=0.0001/\sqrt{2}$. 
    As the interaction~$U$ becomes larger one observes 
    the crossover from a single to individual peaks 
    in the conductance.  {\bfseries\sffamily b},
    the $U$-dependence of the integrated charge~$Q^*$,  
    calculated for wide rectangular cycles  
    for which $X_2$ is varied within $[-\infty,\varepsilon^*]$.
    The values of $\varepsilon^*$ are indicated 
    by arrows of the same color in the main panel.
  }
\end{figure}

Within the two-orbital approximation the many body energies are 
$E_n = E_{\tbox{{gated-well}}}(n) + E_{\tbox{{double-well}}}(N{-}n)$, 
where $n{=}0,1,..,N$,
and $E_{\tbox{{gated-well}}}(n)=\varepsilon n + (1/2)U (n{-}1)n$, 
and $E_{\tbox{{double-well}}}(N{-}n)=-(N{-}n) + (1/4)U(N{-}n{-}1)(N{-}n)$.
The location ${\varepsilon_n=-1 + \frac{1}{2}U \times (N{-}3n{+}2)}$
of the ${n\mapsto (n{-}1)}$ crossing is determined 
from the degeneracy condition ${E_{n}-E_{n{-}1}=0}$, 
where ${n{=}1,2,...N}$. The $N$ crossings are distributed 
within ${\varepsilon\in -1+(N{-}1)[-U,U/2]}$.
\hide{
\be{11}
-1-(N{-}1)U \ \ \le \ \ \varepsilon \ \ \le \ \ -1+\frac{1}{2}(N{-}1)U.
\ee
}
The rescaled version of the control variable is 
${\hat{\varepsilon} = (\varepsilon{+}1)/((N{-}1)U)}$,
and its support is ${-1 < \hat{\varepsilon} < 1/2}$.
The distance between the crossings,
while varying the {gated-well} potential $\varepsilon$, is $U$. 
Once we take $\kappa$ into account 
we get {\em avoided} crossings of 
width ${\delta\varepsilon_n = [(N{+}1{-}n)n/2]^{1/2}\kappa}$.   
If $\kappa/UN$ is large these avoided crossings merge 
and eventually we get one mega crossing.

For the purpose of further analysis we 
apply the two-orbital approximation, 
within which the many body Hamiltonian matrix 
is ${\mathcal{H}_{nm}= E_n\delta_{n,m} - \kappa_n\delta_{n,n\pm1}}$,
where $n{=}0,\cdots,N$ and the couplings are defined 
as ${\kappa_n=\langle n{-}1|\mathcal{H}|n\rangle}$. The calculation involves 
the matrix elements of $b_i^{\dag}b_0$, leading to ${\kappa_n = [(N+1-n)n]^{1/2} \,\kappa}$. 
Analogous expression applies to the current operator where $\kappa$ 
is replaced by ${(k_1{-}k_2)/\sqrt{2}}$.  
For large $U$, as $\varepsilon$ is varied, 
we encounter (say for $N=3$) a sequence of distinct Landau-Zener transitions 
(${|3\rangle \mapsto |2\rangle \mapsto |1\rangle \mapsto |0\rangle}$).
The distance between avoided crossings 
is of order $U$ while their width 
is $\delta \varepsilon_n = \kappa_n$.
The widest crossings are at the center with ${\delta \varepsilon_n \sim N\kappa}$. 
This should be contrasted with 
the energy scales $U$ and $NU$ 
that describe the span of the crossings.  
Accordingly we deduce that for repulsive 
interaction there are three distinct regimes: 
for ${U \ll \kappa/N}$ we have a ``mega crossing";   
for ${U \gg N\kappa}$ we have the ``sequential crossing regime", 
while in the intermediate regime we have a ``gradual crossing".
\hide{
\be{14}
U \ll \kappa & \ \ \ \ & \mbox{mega crossing regime} \nonumber\\
U \gg N\kappa & \ \ \ \ & \mbox{sequential crossing regime}\nonumber
\ee 
}
\hide{
We observe that the regime of behavior depends on 
the ratio $\kappa/U$. If $N$ is not too large 
one can resolve a sequence of two-level crossings.}
Below we summarize the 
results in the various regimes. In particular 
Eq.(\ref{e6}) is obtained from Eq.(\ref{e8}) 
with a two level approximation for each crossing. 
We also related briefly to the ${U<0}$ regime. \\


\begin{figure*}[t]
  \includegraphics[angle=0,width=\hsize,clip,keepaspectratio]{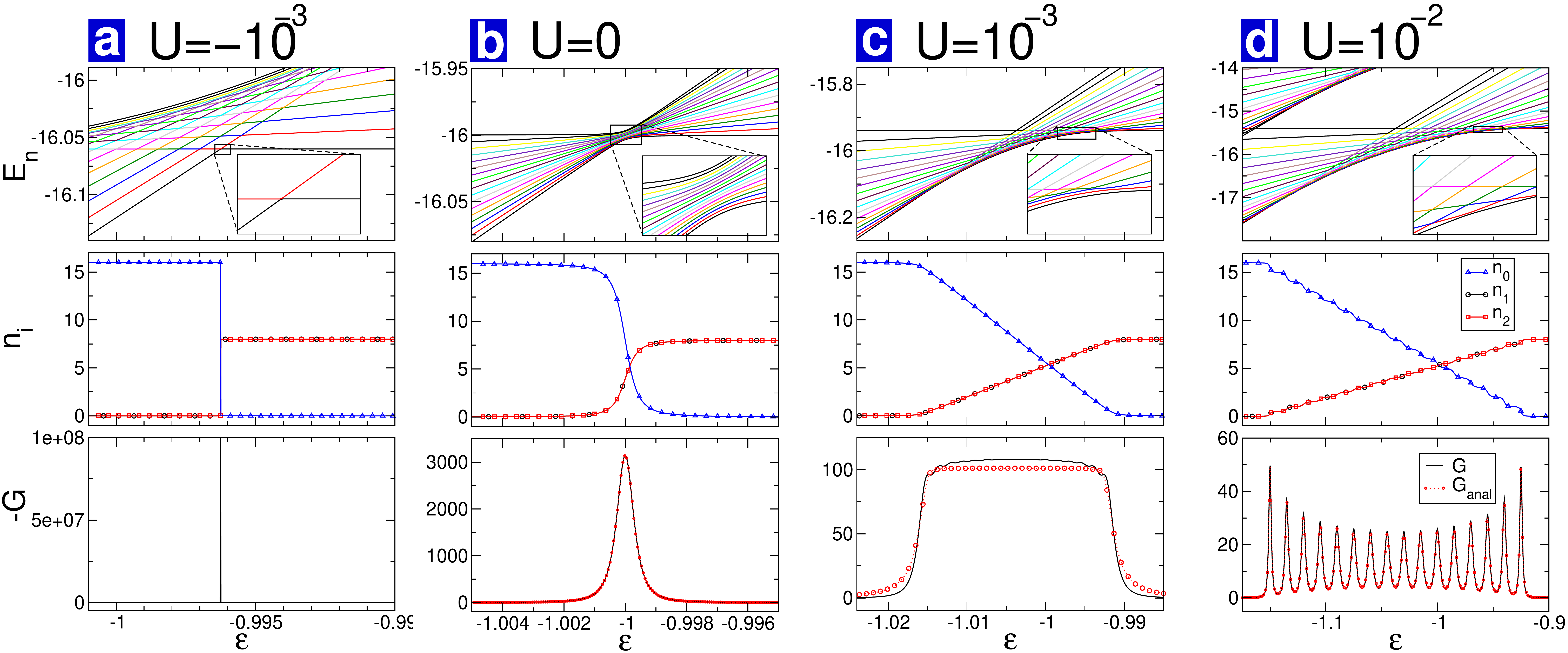}
  \caption{(Color online) 
    {\bfseries\sffamily Evolution of the energy levels, the site
      occupation and the conductance.}
    Further details relating to the data of Fig.~3. 
    We refer to four representative values of~$U$,   
    which are indicated on top of each set of panels.
    {\em Upper panels}:
    the lowest $N{+}1$ energy levels $E_n$ 
    which dominate the conductance $G_2$ are plotted
    as a function of $X_2=\varepsilon$. The insets represent
    magnifications of the indicated areas.
    {\em Middle panels}:
    the site occupations 
    $n_0\;(\mbox{blue }\triangle),\;n_1 (\mbox{black }\circ), n_2\;(\mbox{red }\Box)$.
    {\em Lower panels}:
    the corresponding conductance $G_2$ as a function of $\varepsilon$. 
    Numerical results are represented by solid black lines while the dotted red
    line corresponds to the analytical result (\ref{e4}) in {\bfseries\sffamily b}
    and to (\ref {e6}) in~{\bfseries\sffamily c},~{\bfseries\sffamily d}.
  }
\end{figure*}

{\bf Results --}
{Most of the contribution to the line integral 
in Eq.(\ref{e3}) comes as we change~$X_2$ during the avoided 
crossings that have been discussed in the previous section.
If we close the pumping cycle outside of this limited range,  
then the $X_1$-variation can be safely neglected.  
Accordingly we refer from now on to $G_2=G$ only.} 
An overview of the numerical results for the conductance is shown in Fig.~3, 
where we plot~$G$ as a function of~$X_2$ for various interaction strengths~$U$.
{In the same figure we report the normalized amount of particles $Q$
for various driving cycles. As the }{shape of $G_2$ changes, the dependence of
$Q$ on the $X_2$-span of the pumping cycle becomes of importance. 
Thus, by measuring~$Q$ we obtain information on the strength 
of the interatomic interactions.}

In Fig.~4 more details are presented: besides~$G$ we also plot   
the $X_2$-dependence of the energy levels, and of the site population.    
Four representative values of~$U$ are considered including also the ${U<0}$ case.
Let us discuss the observed results. 
For ${U{=}0}$ all the particles 
cross ``together" from the {gated-well} orbital to the $\varepsilon_{-}$ 
{double-well} orbital. We call this type of dynamics ``mega crossing".
The outcome is just $N$~times the single particle result: 
\be{4}
G = - N \frac{(k_1^2-k_2^2)/2}{[(\varepsilon-\varepsilon_{-})^2+2(k_1{+}k_2)^2]^{3/2}},
\ee
which can be expressed in terms of the control parameters ${(X_1,X_2)}$. 
This result approximately hold as long as ${U \ll \kappa/N}$.
Integrating over a full cycle one obtains  
\be{5}
Q = N\frac{[1+(\kappa R)^2]^{1/2}-1}{\kappa R},
\ee
where $R$ is the radius of the pumping cycle (see Fig.~2).
For small cycles we get ${Q \approx N\kappa R/2}$, 
while for large cycles we get the limiting value ${Q \approx N}$.   
In the other extreme, for very repulsive interaction ($U{\gg}N\kappa$) we get    
\be{6}
G = - \left(\frac{k_1-k_2}{k_1+k_2}\right) 
\sum_{n=1}^N  
\frac{(\delta \varepsilon_n)^2}{[(\varepsilon-\varepsilon_n)^2+(2\delta \varepsilon_n)^2]^{3/2}}.
\ee
For intermediate values ${U\in[\kappa/N,N\kappa]}$,
we find neither the sequential crossing of Eq.(\ref{e6}), 
nor the mega-crossing of Eq.(\ref{e4}), 
but rather a gradual crossing.  
Namely, in this regime, over a range $\Delta X_2=(3/2)(N{-}1)U$ 
we get a constant geometric conductance: 
\be{0}
G \ \ \approx \ \ -\left[\frac{k_1-k_2}{k_1+k_2}\right]\frac{1}{3U}
\ee
which reflects in a simple way the interaction strength. 
This formula was deduced by extrapolating Eq.\ref{e6},  
and then was validated numerically (lower panel of Fig.~4c).

As discussed above for large positive~$U$ 
the {$(N+1)$}-fold ``degeneracy"  of 
the $U{=}0$ Landau-Zener crossing is lifted, 
and we get a sequence of $N$~Landau-Zener crossings  
(for schematic illustration see the lower panel of Fig.~2, 
and compare with the numerical results in the upper panels of Fig.~4). 
Also for $U{<}0$ this {$(N{+}1)$-fold} ``degeneracy" is lifted, 
but in a different way: the levels separate in the ``vertical" (energy) direction 
rather than ``horizontally" (see upper panels of Fig.~4).
Accordingly all the particles execute a single two-level transition 
from the {gated-well} to the {double-well} (see Fig.~4a). 
{This direct Landau-Zener transition from the $n{=}N$ level  
to the $n{=}0$ level is very sharp because it is mediated 
by an $N$-th order virtual transition via the intermediate $n$~states.}  
Accordingly, for sufficiently strong attractive interaction   
all the particles {move together} from the {gated-well} to
{\em one} of the {double-well} sites. When the sign of $X_1$ is reversed they are
transported from one end of the {double-well} to the other end (not shown). 
This should be clearly distinguished from the {$(N+1)$}-fold degenerated transition 
to the lower {double-well} {\em level} which is observed in the $U{=}0$ case.  \\


{\bf Summary --} 
The theoretical \cite{SAZ07,JM06,niu} and experimental \cite{raizen} 
study of driven dynamics in single and double site systems 
is the state of the art. Study of three-site systems adds 
the exciting topological aspect: controlled atomic current 
can be induced using optical lattice technology \cite{AOC05}. 
The actual measurement of induced neutral currents poses a challenge to 
experimentalists. In fact there is a variety of techniques that have been proposed 
for this purpose. For example one can exploit the Doppler effect in the perpendicular
direction, which is known as the rotational frequency shift \cite{doppler}.
The analysis of the prototype trimer system reveals the crucial importance of 
interactions. The interactions are not merely a perturbation: rather they determine 
the nature of the transport process. We expect the induced circulating atomic current
to be extremely accurate, which would open the way to various applications, either as
a new metrological standard, or as a component of a new type of quantum information 
or processing device. 
\vspace{-0.6cm}
\acknowledgements{This research was supported by a grant from the United States-Israel
Binational Science Foundation (BSF) and the DFG (Forschergruppe 760).}

\end{document}